\documentclass[epjCONF,columns]{svjour}

\usepackage{graphics}
\usepackage[latin1]{inputenc}
\session-title{Hadron Collider Physics Symposium 2011}
\begin{document}
\title{QCD studies in the forward region @ LHCb}
\author{Victor Coco\thanks{\email{victor.coco@cern.ch}}, \textit{on behalf of LHCb collaboration} }
\institute{Nikhef, Amsterdam, The Netherlands}
\abstract{
The LHCb experiment at the LHC is fully instrumented over a unique pseudorapidity range in the forward region. Although it has been designed for $b$-physics, LHCb is able to provide valuable informations on particle production in this region of phase space. 
Therefore QCD studies have been performed with the LHCb detector on $pp$ collisions at $\sqrt{s}=900~\mathrm{GeV}$ and $\sqrt{s}=7~\mathrm{TeV}$. The measurement of charged particles multiplicity at $\sqrt{s}=7~\mathrm{TeV}$, $\bar{\Lambda}/\Lambda$, $\bar{\Lambda}/K_{s}^{0}$ production ratios at $\sqrt{s}=900~\mathrm{GeV}$ and $\sqrt{s}=7~\mathrm{TeV}$, as well as light hadrons ($p$,$K$,$pi$) production ratios at $\sqrt{s}=900~\mathrm{GeV}$ and $\sqrt{s}=7~\mathrm{TeV}$ are reported.
} 
\maketitle
\section{Introduction}
\label{intro}

The LHCb experiment is dedicated to CP violation and rare decay measurements involving b and c hadrons at the LHC~\cite{lhcb}. Since most of the correlated $b\bar{b}$ pair are produced at small angle with respect to the beam line, the LHCb detector has been designed as a single arm spectrometer instrumented over the $2<\eta<5$ pseudo-rapidity range. This unique coverage of the forward region, together with excellent tracking and particle identification performances allows the probing particle production at the LHC in an uncharted region of phase space.

The forward hadron production measurements gives important inputs to tune the hadronisation models in the LHCb acceptance, since these models have been extrapolated from measurements performed in a different energy and rapidity regime.
Production cross-section of several particles have been measured at LHCb. $K_{s}^{0}$ cross section measurement at $\sqrt{s}=900~\mathrm{GeV}$~\cite{ks} and $\phi$ cross section measurement at $\sqrt{s}=7~\mathrm{TeV}$~\cite{phi} probe strangeness production. $J/\psi$~\cite{jpsi}, $\psi(2S)$~\cite{psi}, $\Upsilon$~\cite{upsilon}, prompt charm~\cite{c} and $b$~\cite{b} cross section measurements at $\sqrt{s}=7~\mathrm{TeV}$ probe heavy flavour production. The measurements of charged particles multiplicity at $\sqrt{s}=7~\mathrm{TeV}$~\cite{chmult}, as well as $V_0$ production ratios: $\bar{\Lambda}/\Lambda$,$\bar{\Lambda}/K_{s}^{0}$~\cite{v0} and light hadrons ($\pi$,$K$,$p$) production ratio~\cite{p} both at $\sqrt{s}=900~\mathrm{GeV}$  and $\sqrt{s}=7~\mathrm{TeV}$ are presented in the following.

\section{Charged particles multiplicity at $\sqrt{s}=7~\mathrm{TeV}$}
\label{sec:chpartmult}
The measurement of charged particles multiplicity at the LHCb experiment\cite{chmult}, provides input for modelling the underlying event structure in high energy $pp$ collisions in the forward region. 

The charged particles are reconstructed in the vertex locator (VELO) detector situated close to the interaction region. It provides a high reconstruction efficiency and purity for charged particles in the $\eta$ range $-2.5<\eta<-2.0$ and $2.0<\eta<4.5$. Since the magnetic field in the VELO is negligible, no momentum information is to VELO-reconstructed charged particles. In order to probe more energetic collision, the charged particles multiplicity is also measured in events where at least one track in the pseudo-rapidity range of $2.5<\eta<4.5$ went through the whole tracking system, which allow the measurement of its momentum. This track is required to have a transverse momentum greater than $1~GeV/c$.  

\begin{figure}[b]
\resizebox{0.49\columnwidth}{!}{\includegraphics{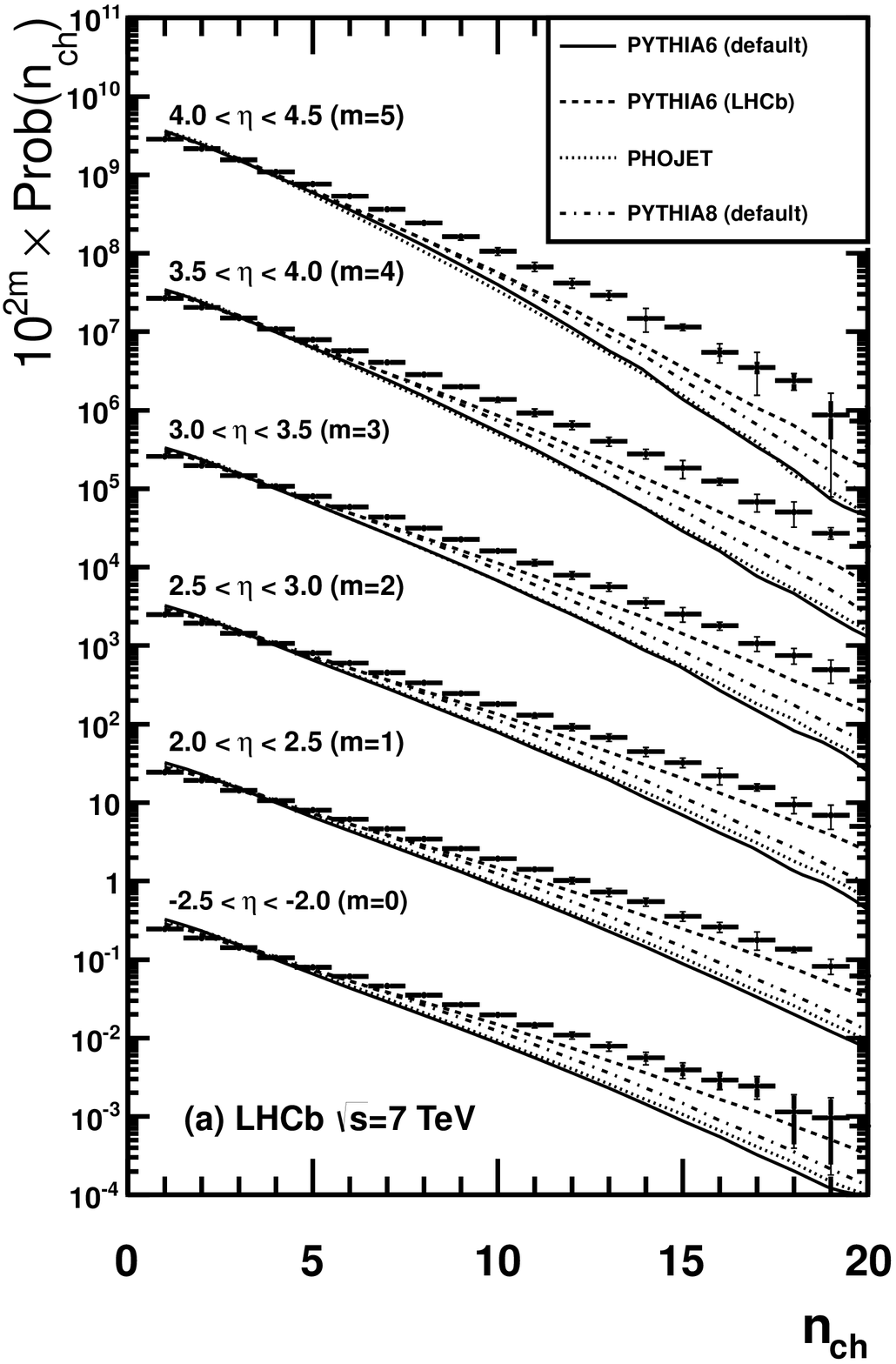}}
\resizebox{0.49\columnwidth}{!}{\includegraphics{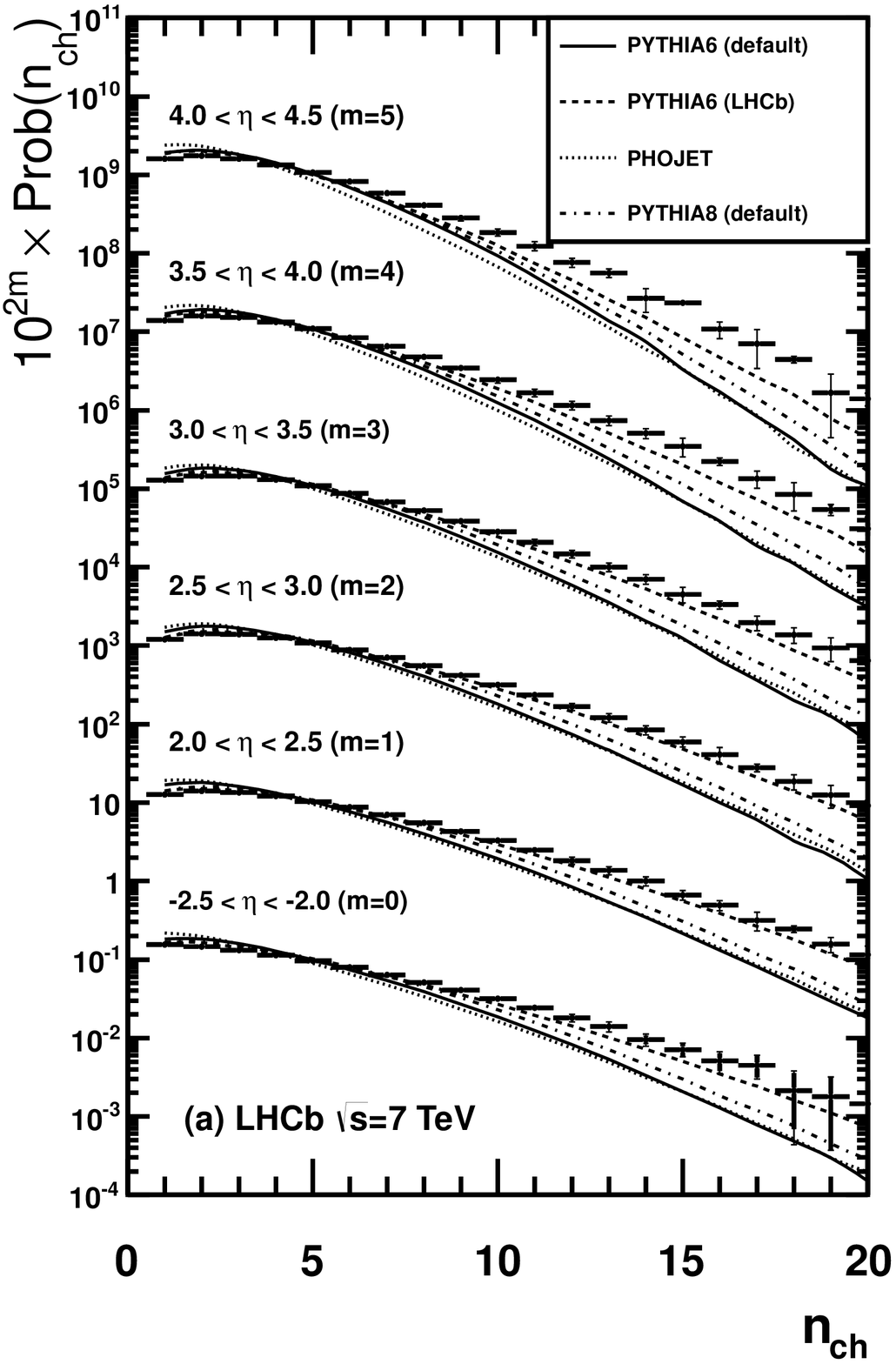}}
\caption{\label{fig:chMult} The multiplicity distribution in $\eta$ bins (shown as points with statistical error bars)
with predictions of different event generators. The inner error bar represents the statistical
uncertainty and the outer error bar represents the systematic and statistical uncertainty on the
measurements. Predictions are from Pythia 6, Phojet and Pythia 8, comparison with other predictions can be found in~\cite{chmult}.
Left is for all events, right is for events with at least one track with $p_T>1~GeV/c$ in $2.5<\eta<4.5$.}
\end{figure}

The analysis is performed on $1.5$ million events for each magnet polarity, which were triggered by requiring at least one reconstructed VELO track. This dataset has low pile up, with only $3.7\pm 0.4\%$ of events having more than one interaction. 
Monte Carlo simulation is used to correct for acceptance, resolution effects and to estimate the secondary charged particles contamination. The main source of systematic uncertainty is due to the tracking efficiency determination and is estimated to be $4\%$ overall. Other source of systematics uncertainty are small compared to it. The event multiplicity is obtained by unfolding of the migration due to reconstruction inefficiencies. 

Figure~\ref{fig:chMult} left, shows the unfolded multiplicity distribution in $\eta$ bins for all events, while figure~\ref{fig:chMult} right shows it for harder events with at least one track with $p_T>1~GeV/c$ in $2.5<\eta<4.5$.
Several event generator have been compared with the data. None are fully able to describe the multiplicity distributions, even though the agreement is better for hard QCD events.

\section{$\bar{\Lambda}/\Lambda$ and $\bar{\Lambda}/K_{s}^{0}$ ratios}
\label{sec:lambdaksratio}

The production ratios of baryon/anti-baryon, such as $\bar{\Lambda}/\Lambda$, allow to study the baryon-number transport from the beam particles to the final state. $\bar{\Lambda}/K_{s}^{0}$ ratio on the other hand is a measure of the baryon to meson suppression, which is a good test for different fragmentation models.

High purity samples of prompt $K_{s}^{0}$ decaying into $\pi^{+}\pi^{-}$ and $\Lambda$, $\bar{\Lambda}$ decaying into $p\pi$ are selected based on a Fisher discriminant combining the impact parameter of the $V^0$ particles and their daughters. Only events with one primary vertex are selected to avoid diffractive event contribution. No particle identification is used here~\cite{v0}.

\begin{figure}[b]
\begin{center}
\resizebox{0.75\columnwidth}{!}{\includegraphics{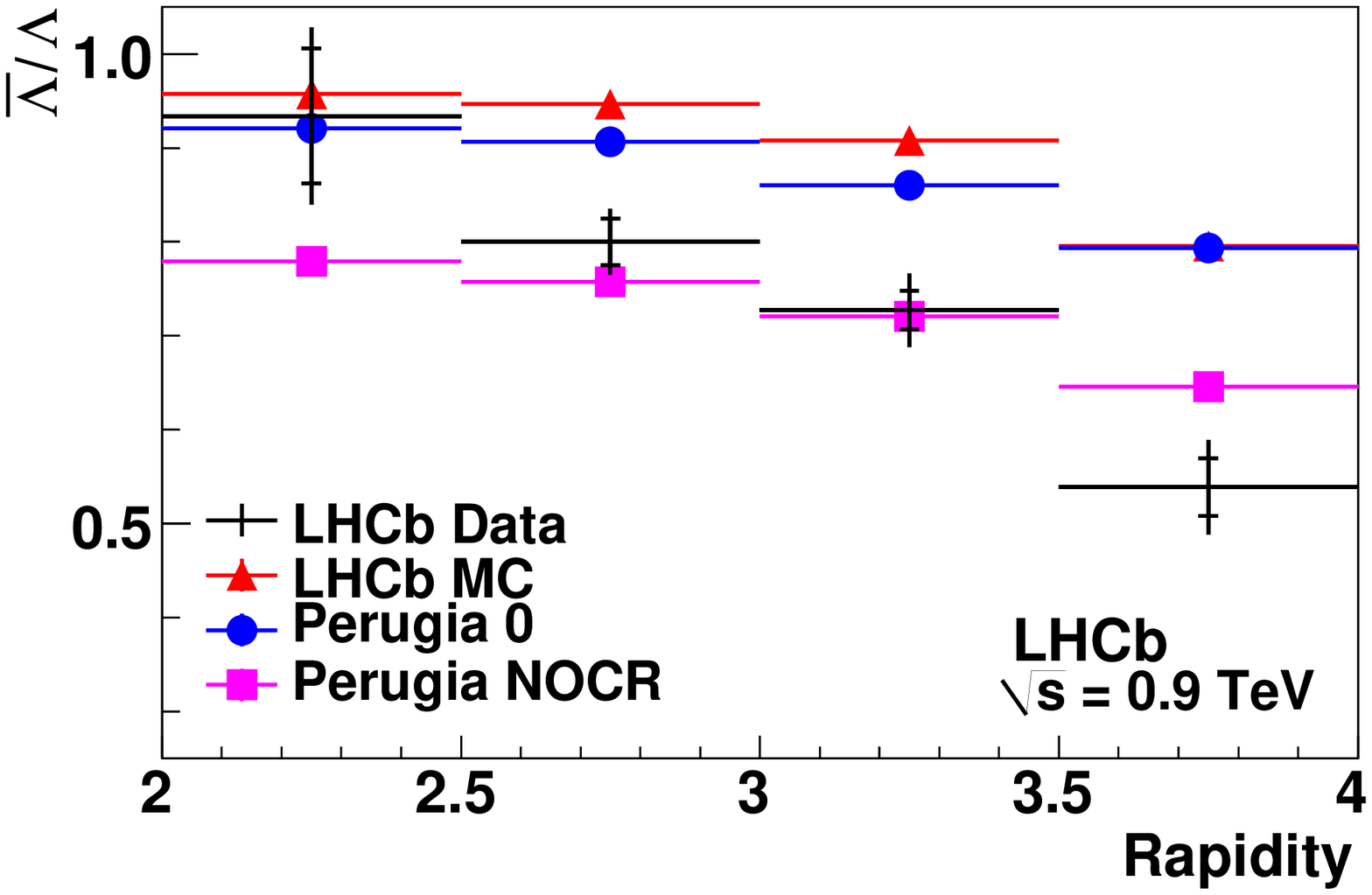}}
\resizebox{0.75\columnwidth}{!}{\includegraphics{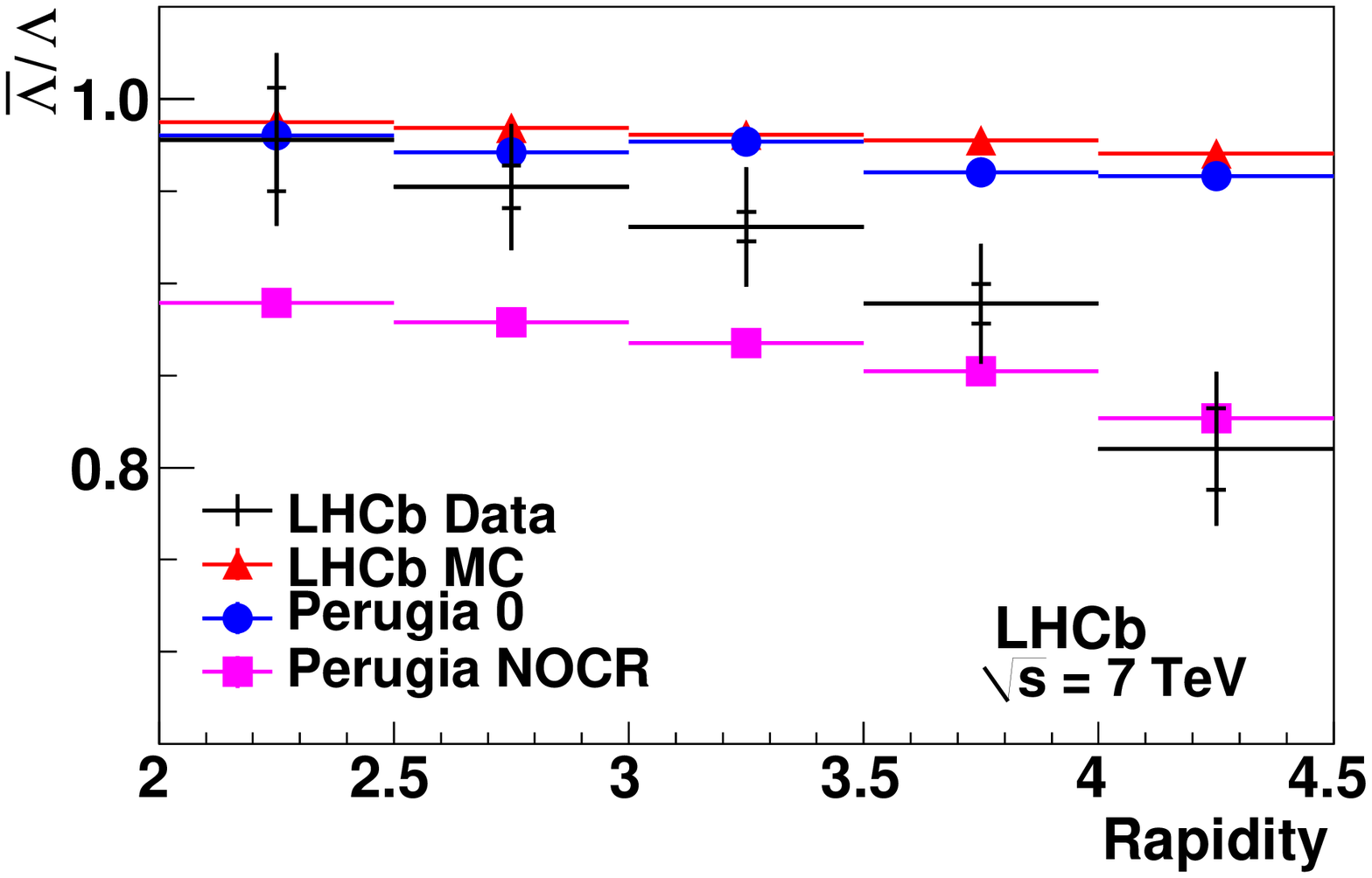}}
\end{center}
\caption{\label{fig:lambdaRatio} The production cross-section ratio $\bar{\Lambda}/\Lambda$ at $\sqrt{s}=900~\mathrm{GeV}$ and $\sqrt{s}=7~\mathrm{TeV}$ as a function of rapidity  compared with the predictions of the LHCb MC tune~\cite{lhcbmc}, Perugia 0 and Perugia NOCR~\cite{perugia}. Vertical lines show the combined statistical and systematic uncertainties and the short horizontal bars (where visible) show the statistical component.}
\end{figure}

\begin{figure}
\begin{center}
\resizebox{0.75\columnwidth}{!}{\includegraphics{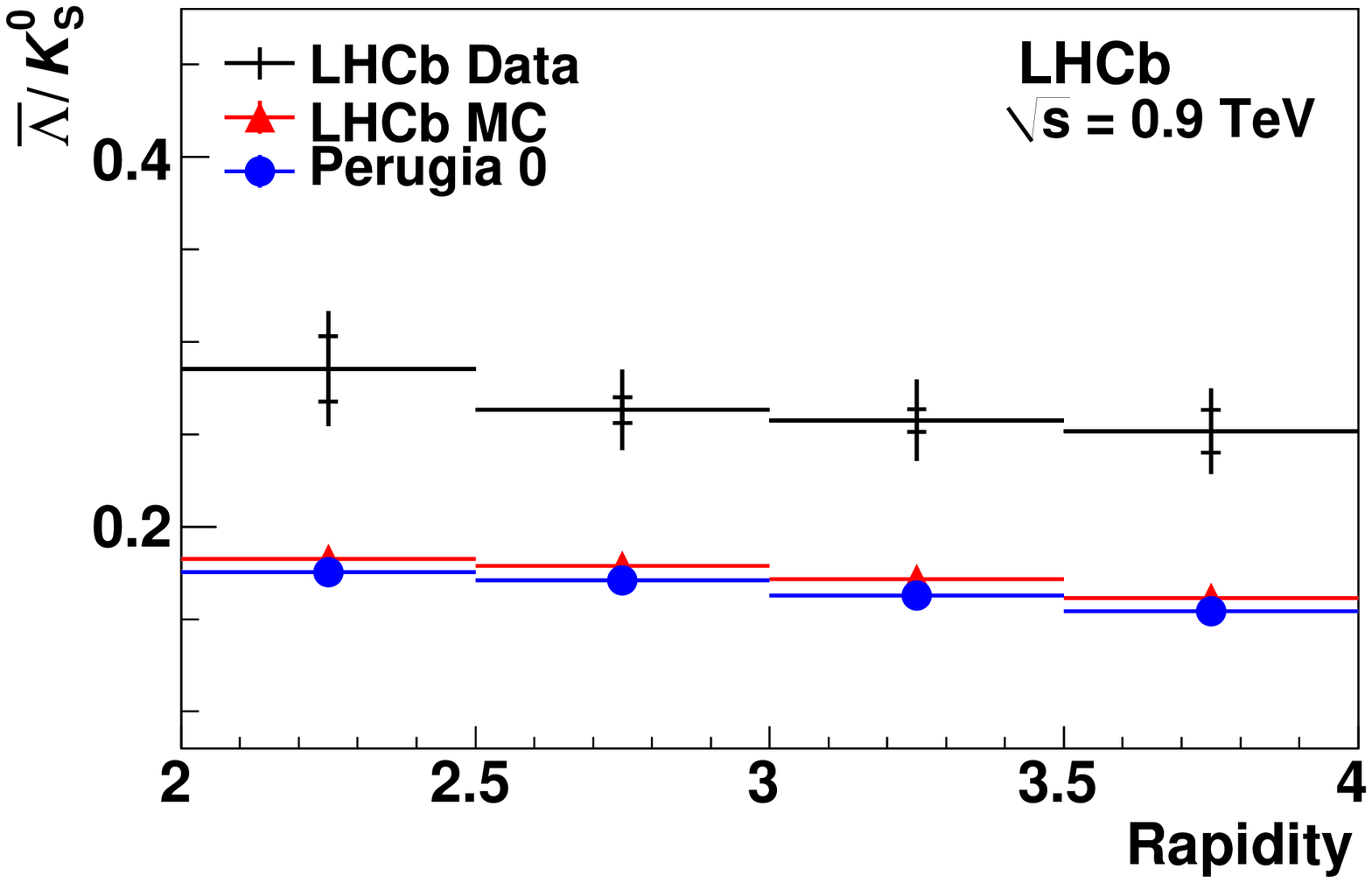}}
\resizebox{0.75\columnwidth}{!}{\includegraphics{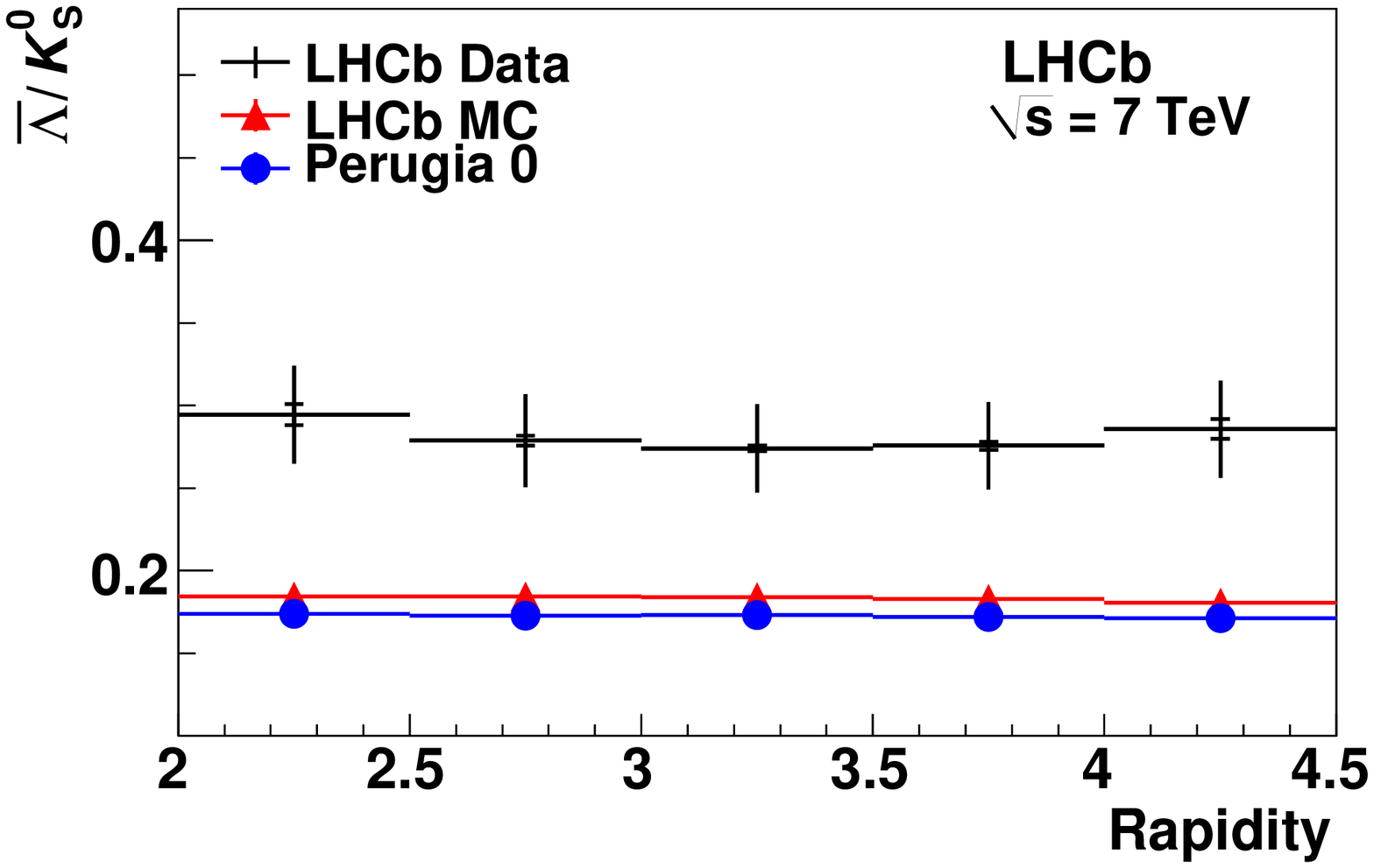}}
\end{center}
\caption{\label{fig:lambdaKsRatio} The production cross-section ratio $\bar{\Lambda}/K_{s}^{0}$ at $\sqrt{s}=900~\mathrm{GeV}$ and $\sqrt{s}=7~\mathrm{TeV}$ as a function of rapidity compared with the predictions of the LHCb MC tune~\cite{lhcbmc} and Perugia 0~\cite{perugia}. Vertical lines show the combined statistical and systematic uncertainties and the short horizontal bars (where visible) show the statistical component.}
\end{figure}

\begin{figure}[b]
\begin{center}
\resizebox{0.73\columnwidth}{!}{\includegraphics{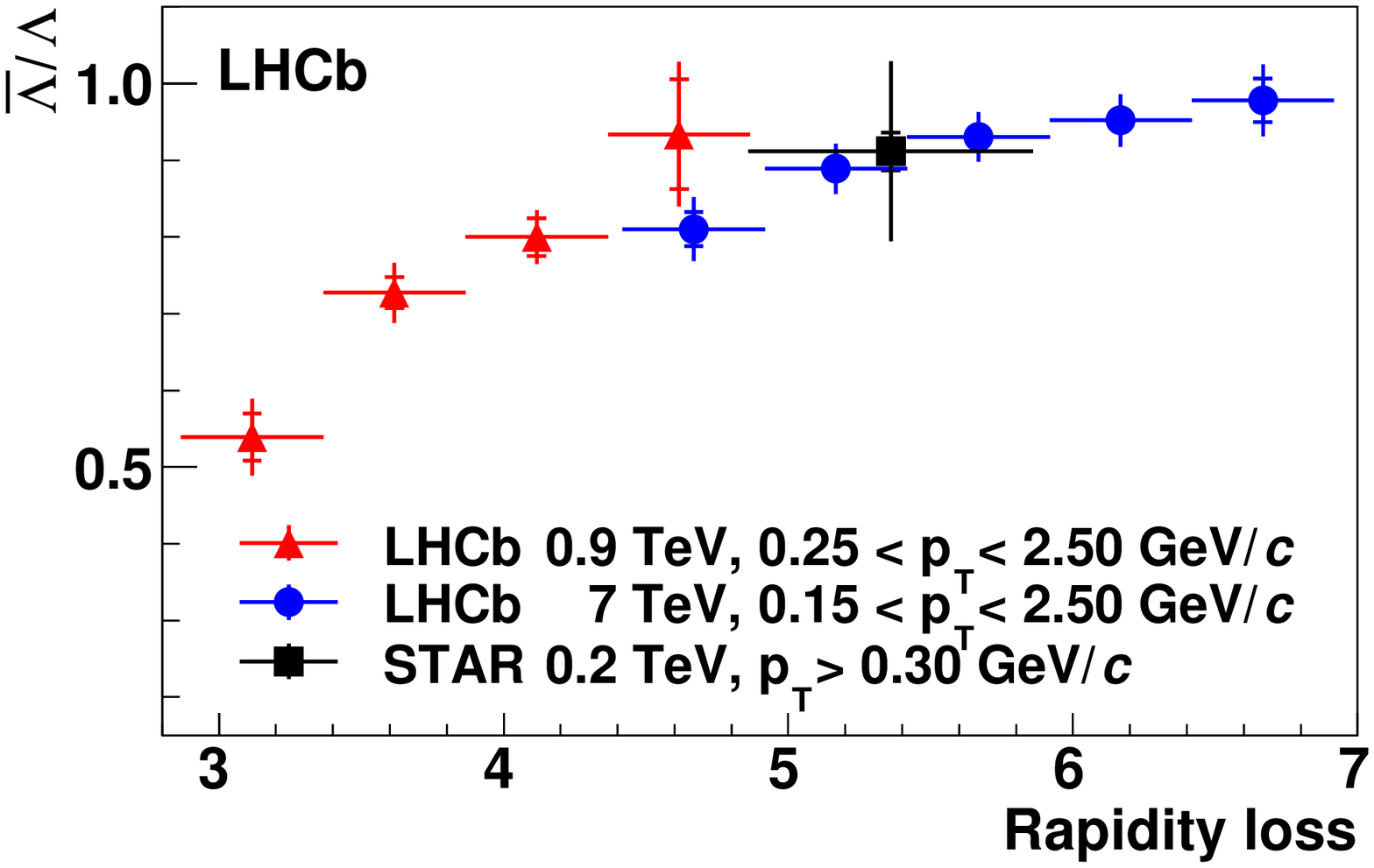}}
\resizebox{0.73\columnwidth}{!}{\includegraphics{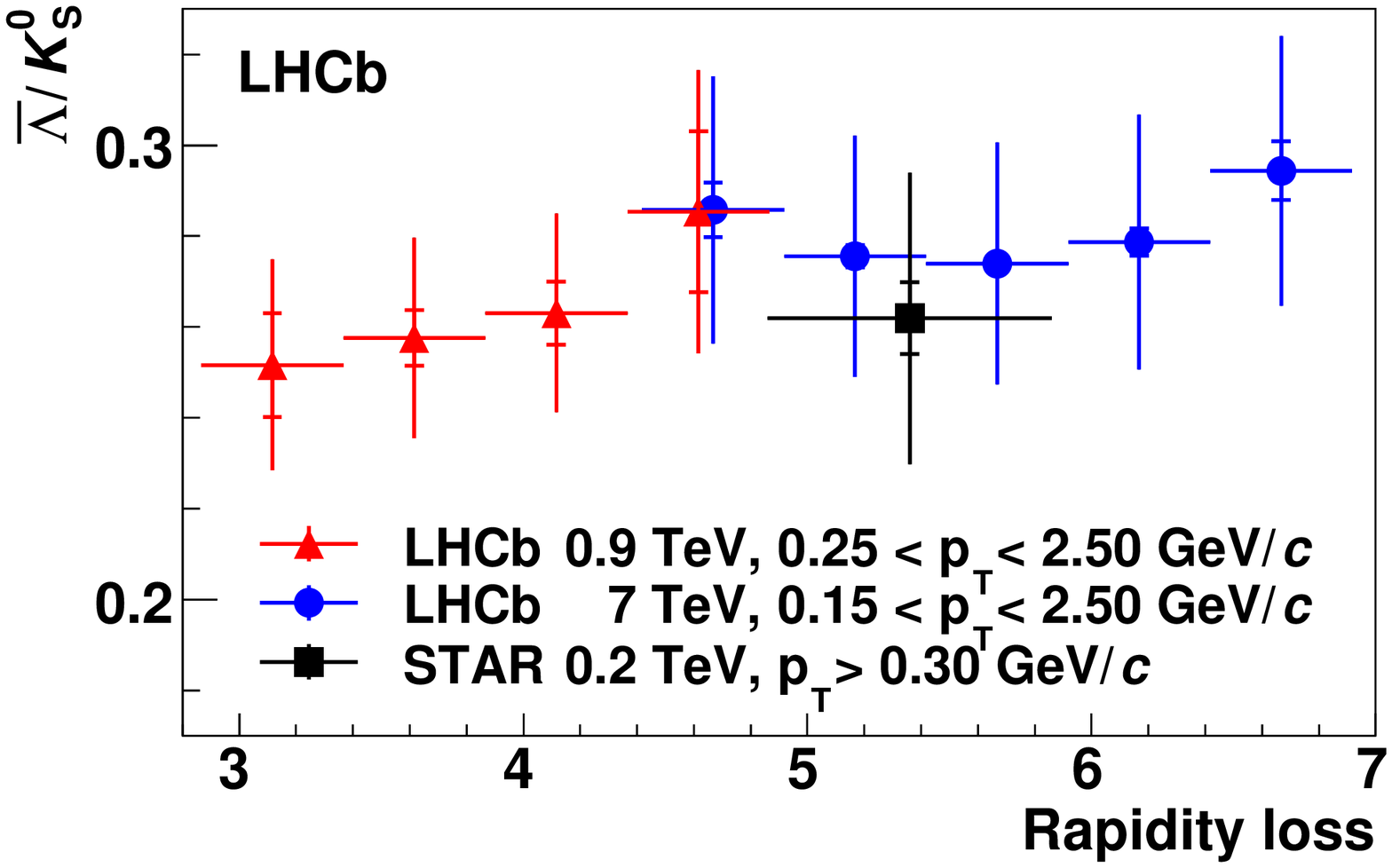}}
\caption{\label{fig:lambdaratioy} The ratios  $\bar{\bar{\Lambda}}/\Lambda$ (top) and $\bar{\Lambda}/K_{s}^{0}$ (bottom) from LHCb are compared at both at $\sqrt{s}=900~\mathrm{GeV}$ (red) and $\sqrt{s}=7~\mathrm{TeV}$ (blue) with the published results from STAR~\cite{star} (black) as a function of rapidity loss. Vertical lines show the combined statistical and systematic uncertainties and the short
horizontal bars (where visible) show the statistical component.}
\end{center}
\end{figure}

The systematic uncertainty is reduced since the major uncertainty on the production cross-sections, that comes from absolute luminosity measurement, cancels through the ratio. The remaining sources of systematic uncertainty come from kinematic correction of the Monte Carlo simulation used to evaluate selection efficiency, the uncertainty in the interaction with material and the diffractive events pollution. Depending on the bins in $\eta$ and $p_T$, they add up to 0.02-0.06 for $\bar{\Lambda}/\Lambda$ and 0.02-0.03 for $\bar{\Lambda}/K_{s}^{0}$.

The measurement is performed with $0.3~\mathrm{nb^{-1}}$ at $\sqrt{s}=900~\mathrm{GeV}$ and $1.38~\mathrm{nb^{-1}}$ at $\sqrt{s}=7~\mathrm{TeV}$.
Figure~\ref{fig:lambdaRatio} shows that $\bar{\Lambda}/\Lambda$ is in rather good agreement with Perugia 0 tune for Pythia at low rapidity while at high rapidity, extreme models of baryon transport seams to be favoured~\cite{perugia}. This behaviour is observable both at $\sqrt{s}=900~\mathrm{GeV}$ and $\sqrt{s}=7~\mathrm{TeV}$.
In Figure~\ref{fig:lambdaKsRatio}, $\bar{\Lambda}/K_{s}^{0}$ shows an excess over the whole range of rapidity, at both center of mass energy.

Another way to present the baryon number transport results is to show the production ratio of anti-baryon to baryon as a function of the rapidity loss, $\Delta y = y_{beam}-y$ where $y_{beam}$ is the rapidity of the incoming beam, Figure~\ref{fig:lambdaratioy}. It allows to compare the LHCb results with the measurements of the previous experiments. It also shows that there is no significant energy scale violation between the results at $\sqrt{s}=900~\mathrm{GeV}$ and $\sqrt{s}=7~\mathrm{TeV}$.

\section{Light hadrons production ratios}
\label{sec:lighthadronratio}
\begin{figure}[!b]
\begin{center}

\resizebox{0.8\columnwidth}{!}{\includegraphics{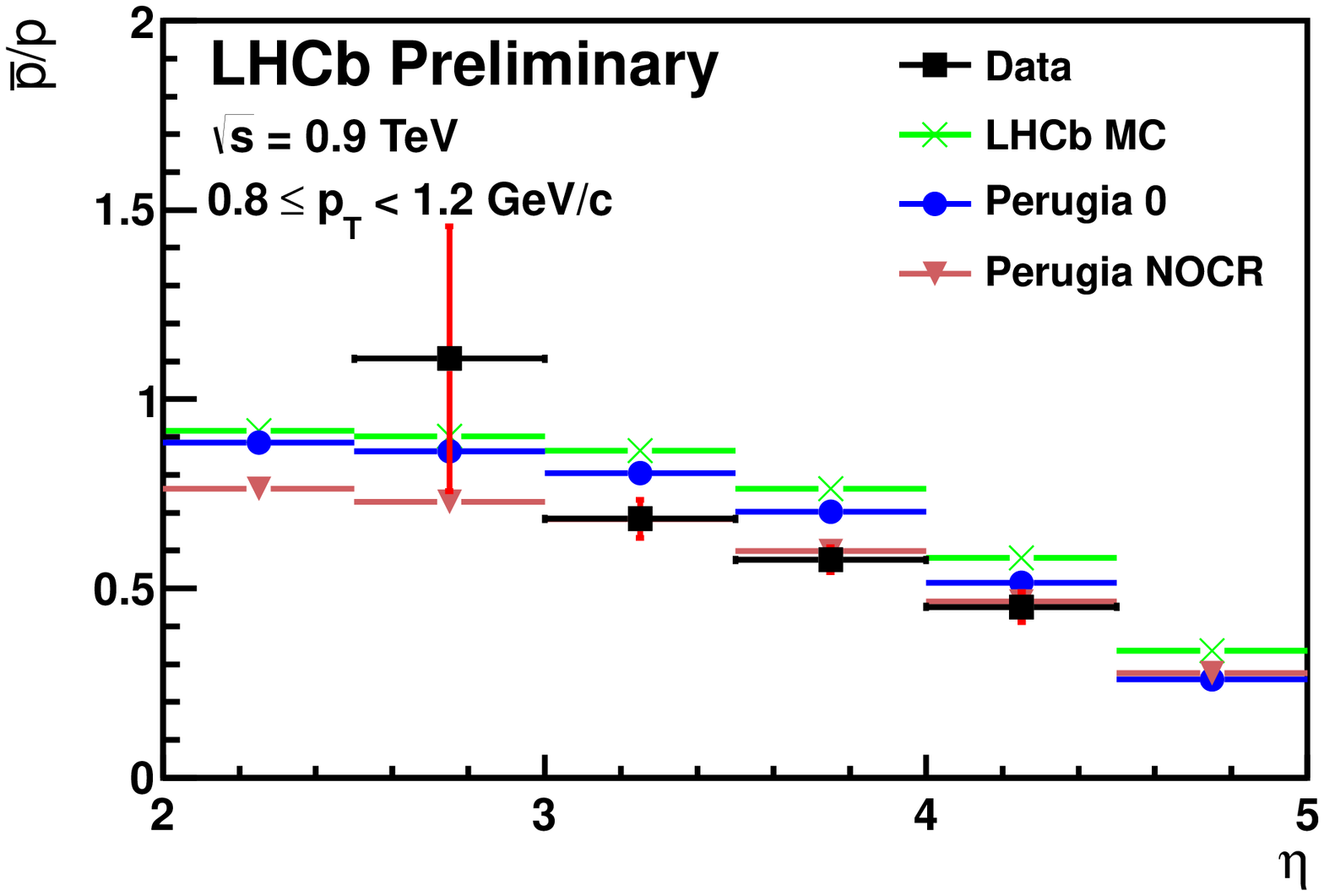}}
\resizebox{0.8\columnwidth}{!}{\includegraphics{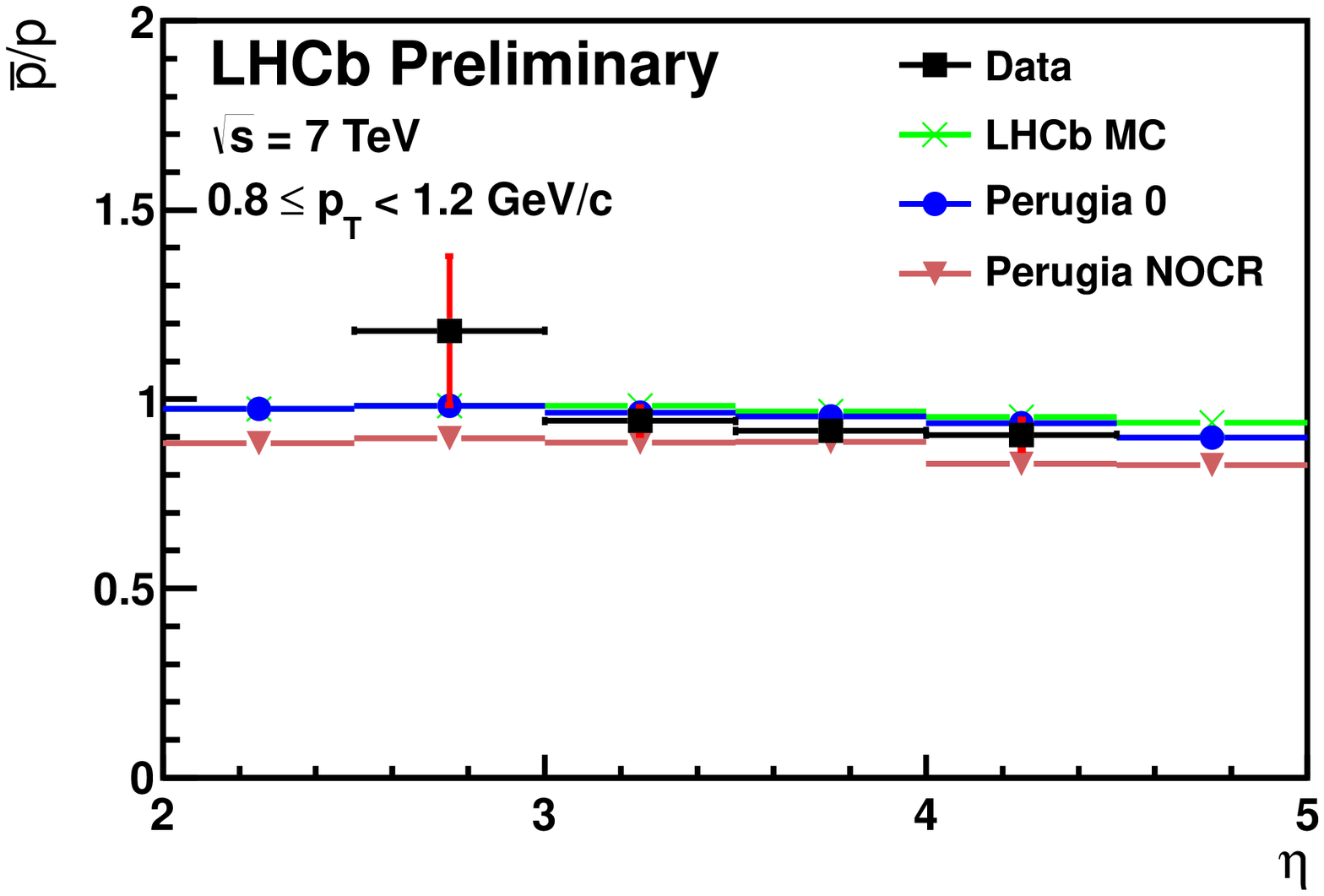}}
\end{center}
\caption{\label{fig:ppratio}Distribution of the $\bar{p}/p$ production ratio against rapidity, $0.8<p_T<1.2~\mathrm{GeV/c}$. Top for $\sqrt{s}=900~\mathrm{GeV}$, bottom for $\sqrt{s}=7\mathrm{TeV}$. Data are compared with the predictions of the LHCb MC tune~\cite{lhcbmc} and Perugia 0 and Preugia NOCR tunes~\cite{perugia}}.
\end{figure}

The measurement of light hadron production ratios at LHC is another input that can be used to validate hadronisation models. In addition $\bar{p}/p$ production ratio probes the baryon number transport. The production ratios $\bar{p}/p$, $K^{-}/K^{+}$, $\pi^{-}/\pi^{+}$, $(\bar{p}+p)/(K^{-}+K^{+})$,$(\bar{p}+p)/(\pi^{-}+\pi^{+})$ and $(K^{-}+K^{+})/(\pi^{-}+\pi^{+})$ have been measured in the forward region, both at $\sqrt{s}=900~\mathrm{GeV}$ and $\sqrt{s}=7~\mathrm{TeV}$ center of mass energy. The preliminary results shown here are an update of~\cite{p}. The measurement is performed  in three bins of transverse momentum, $p_T<0.8~\mathrm{GeV/c}$,$0.8<p_T<1.2~\mathrm{GeV/c}$, and $p_T>1.2~\mathrm{GeV/c}$ and five bins of rapidity, in the range $2<y<5$, with $0.3~\mathrm{nb^{-1}}$ at $\sqrt{s}=900~\mathrm{GeV}$ and $1.8~\mathrm{nb^{-1}}$ at $\sqrt{s}=7~\mathrm{TeV}$.

Prompt light hadrons with $p>5~\mathrm{GeV/c}$ are selected in events with at least one reconstructed primary vertex. Exploiting the powerful hadronic separation capabilities of the LHCb RICH system, the cross contamination between species have been evaluated. The performance of the particle identification is calibrated on kinematically isolated samples of $\phi\rightarrow K^+K^-$, $K_s\rightarrow\pi^+\pi^-$ and $\Lambda\rightarrow\pi p$. 
The main systematic comes from cross contamination between $p$, $K$ and $\pi$. It depends on which ratio is considered and varies from the percent level up to tens percent in the extreme region of rapidity.

\begin{figure}
\begin{center}

\resizebox{0.8\columnwidth}{!}{\includegraphics{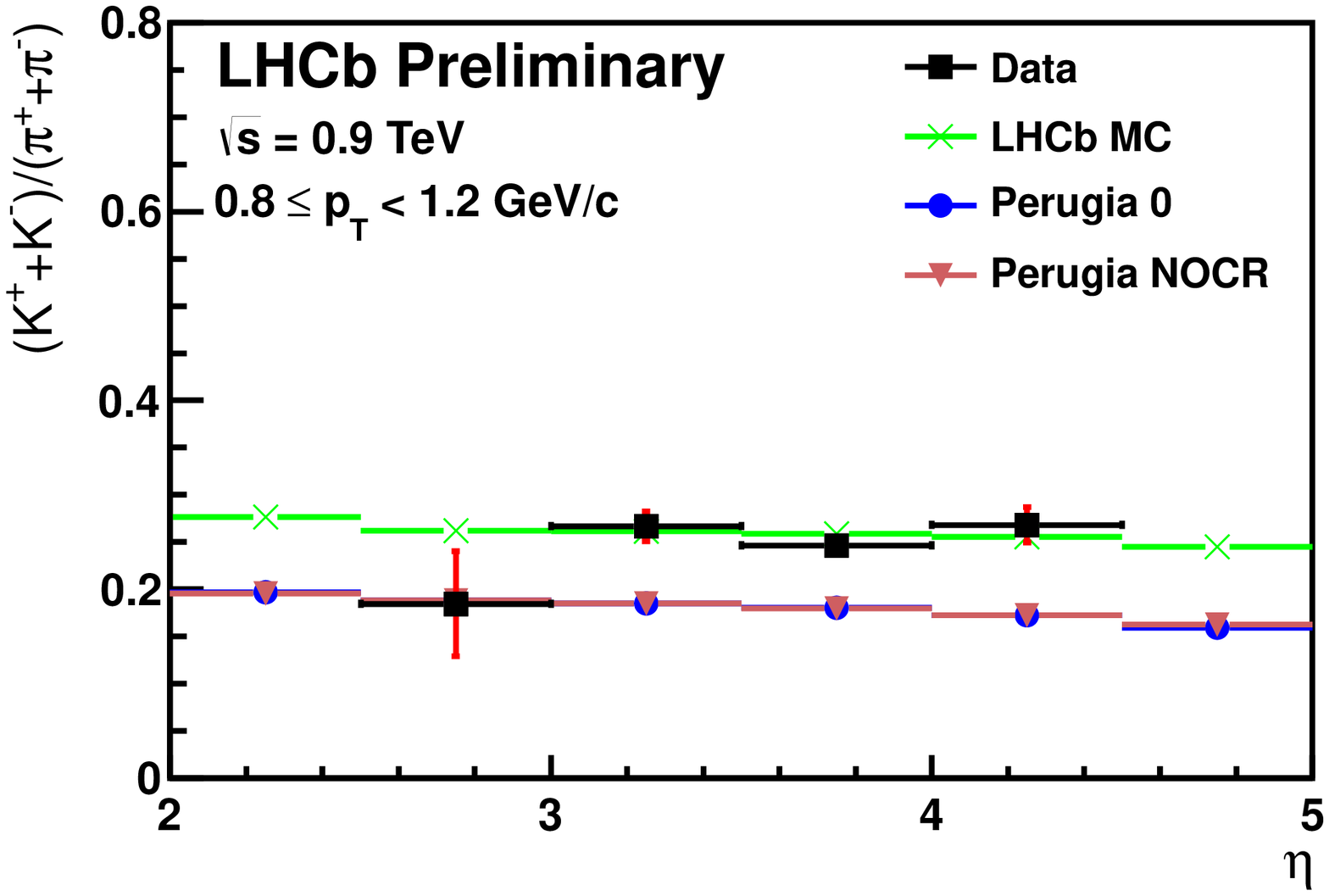}}
\resizebox{0.8\columnwidth}{!}{\includegraphics{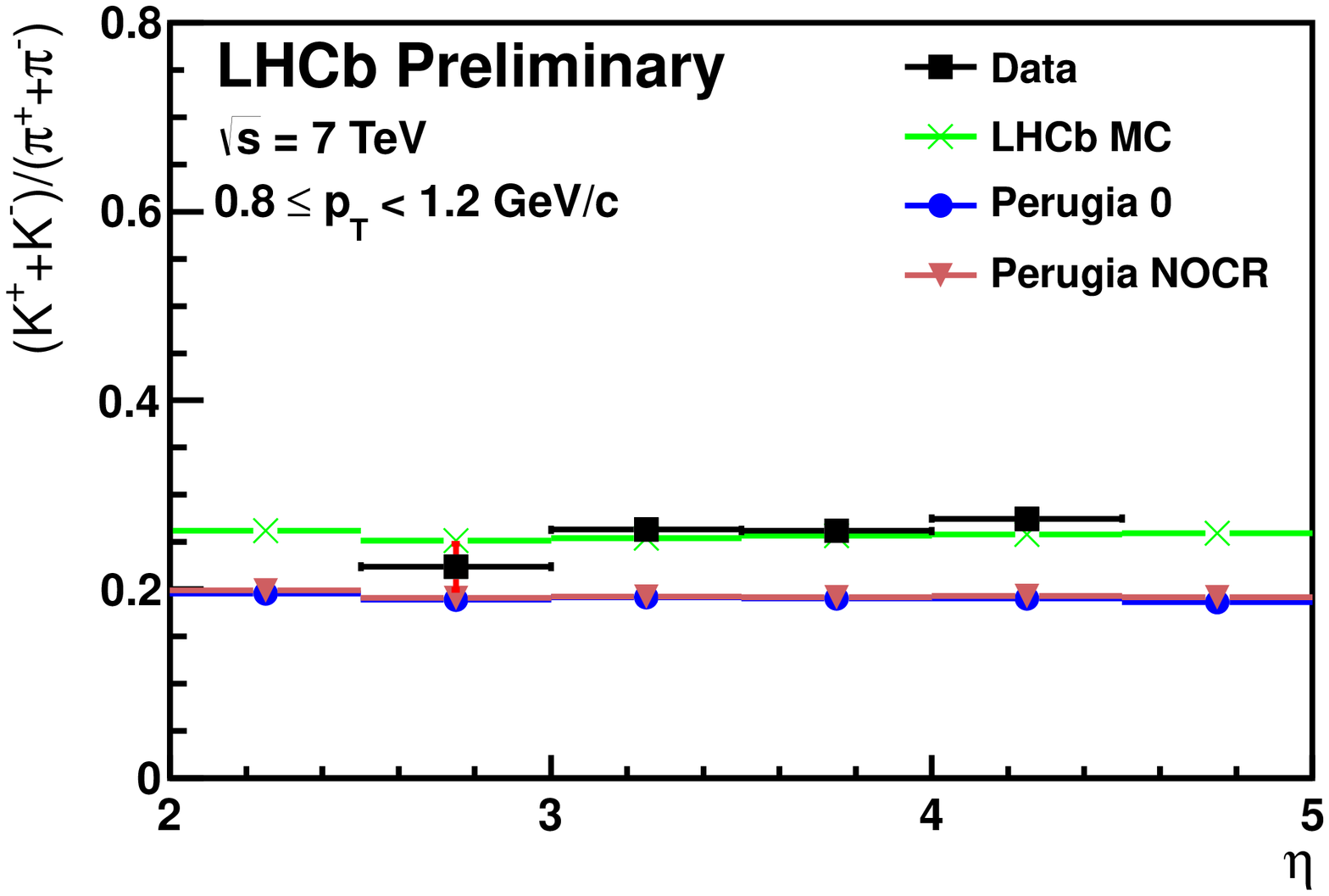}}
\end{center}
\caption{\label{fig:kpiratio}Distribution of the $(K^{-}+K^{+})/(\pi^{-}+\pi^{+})$ production ratio against rapidity, $0.8<p_T<1.2~\mathrm{GeV/c}$. Top for $\sqrt{s}=900~\mathrm{GeV}$, bottom for $\sqrt{s}=7\mathrm{TeV}$. Data are compared with the predictions of the LHCb MC tune~\cite{lhcbmc} and Perugia 0 and Preugia NOCR tunes~\cite{perugia}}.
\end{figure}

Results have been compared with different MC models. If the general behaviour is reproduced by the models, there is still room for improvements. Figure~\ref{fig:ppratio} and figure~\ref{fig:kpiratio} show as illustration the results of the analysis for $\bar{p}/p$ and $(K^{-}+K^{+})/(\pi^{-}+\pi^{+})$, for the $[0.8;12~\mathrm{GeV/c}]$ $p_T$ bin.

\begin{figure}
\begin{center}
\resizebox{0.9\columnwidth}{!}{\includegraphics{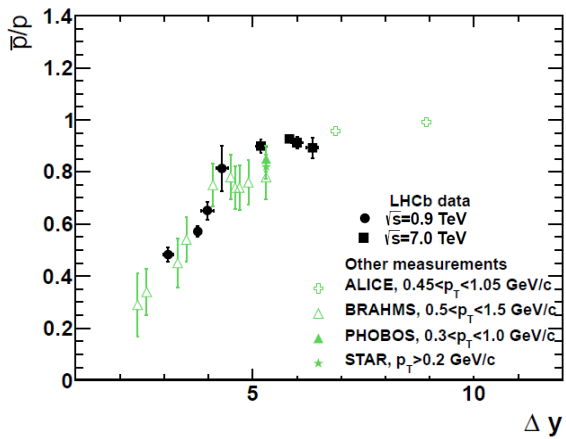}}
\caption{\label{fig:rapLoss} $\Delta y$ from LHCb~\cite{p} and other experiments~\cite{povp}, weighted average in $p_T$. Uncertainties are statistical + systematics, except in the case of data collected at the ISR where only statistical uncertainties are shown.}
\end{center}
\end{figure}

In the following, highlight is put on the $\bar{p}/p$ production ratio since it allows further tests of the baryon number transport. Results are consistent with Monte Carlo models at $\sqrt{s}=7~\mathrm{TeV}$ but differ significantly at $\sqrt{s}=900~\mathrm{GeV}$, especially at low $p_T$, as shown in Figure~\ref{fig:ppratio}.
Like for the $\bar{\Lambda}/\Lambda$ production ratio measurement, $\bar{p}/p$ production ratio is shown as function of the rapidity loss $\Delta y$, Figure~\ref{fig:rapLoss}. The LHCb measurements are compatible with those of the previous experiments~\cite{povp}, but significantly more precise. The ALICE experiment measurement cover the high rapidity loss region with good precision, showing the complementarity at LHC between experiments covering the central region and LHCb for production measurements.

\section{Conclusions}
The charged particle multiplicity measured at LHCb shows that the number of charged particles produced forward is underestimated by most of the Monte Carlo models in the forward region of $pp$ collisions at $\sqrt{s}=7~\mathrm{TeV}$. 
The measurements of $V_0$ production ratios suggest lower strange baryon suppression and higher baryon number transport than in the Monte Carlo models that have been investigated. The LHCb data are consistent with data from lower energy experiments. Light hadron production ratios of $\bar{p}/p$, $K^{-}/K^{+}$, $\pi^{-}/\pi^{+}$, $(\bar{p}+p)/(K^{-}+K^{+})$,$(\bar{p}+p)/(\pi^{-}+\pi^{+})$ and $(K^{-}+K^{+})/(\pi^{-}+\pi^{+})$ have been measured both at $\sqrt{s}=900~\mathrm{GeV}$ and $\sqrt{s}=7~\mathrm{TeV}$, and compared to some expectation from Monte Carlo models. It suggests that improvements are needed to fully reproduce their behaviour as function of rapidity and transverse momentum.

%


\end{document}